\documentclass[aps,prl,reprint,groupedaddress]{revtex4-2}

\usepackage{color}
\usepackage{soul,color}
\usepackage{graphicx}

\begin{document}


\title{Significant loss suppression and large induced chirality via cooperative near- and far-field coupling in plasmonic dimer nanoantennas}


\author{Xiaoqing Luo$^1$}
\author{Rixing Huang$^{1}$}
\author{Dangyuan Lei$^{2}$}
\author{Guangyuan Li$^{1,3}$}
\email[]{gy.li@siat.ac.cn}
\affiliation{$^1$CAS Key Laboratory of Human-Machine Intelligence-Synergy Systems, Shenzhen Institutes of Advanced Technology, Chinese Academy of Sciences, Shenzhen 518055, China}
\affiliation{$^2$Department of Materials Science and Engineering, City University of Hong Kong, Hong Kong 999077, China}
\affiliation{$^3$Shenzhen College of Advanced Technology, University of Chinese Academy of Sciences, Shenzhen, 518055, China}



\date{\today}

\begin{abstract}
Plasmonic nanoantennas containing nano-gaps support ``hotspots'' for greatly enhanced light-matter interactions, but suffer from inherent high losses, a long-standing issue that hinders practical applications. Here we report a strategy to significantly suppress the losses of plasmonic dimer nanoantennas. Specifically, by introducing the {\sl concept} of cooperative near- and far-field coupling, we observed an unprecedented transition from the weak coupling of localized resonances to strong coupling of collective (nonlocal) resonances, showing robustness to the gap distance between the dimer. We develop a generalized lattice sum approximation model to describe this transition and reveal its origins: the off-diagonal element of the anisotropic polarizability tensor due to near-field coupling, and the anisotropic lattice sums due to far-field coupling. This strong coupling leads to loss-suppressed plasmonic resonances with large modulation depths and meanwhile extremely high measured quality factors up to 3120 in the near-infrared regime, exceeding the {\sl record} in the near infrared regime. Additionally, high-$Q$ and large chiroptical responses can also be induced for achiral planar dimers under the critical coupling condition.
This work paves an avenue toward extremely low-loss plasmonic devices, either chiral or not, for diverse important applications. 
\end{abstract}


\maketitle

\begin{figure*}[!hbt]
\centering
\includegraphics[width=170 mm]{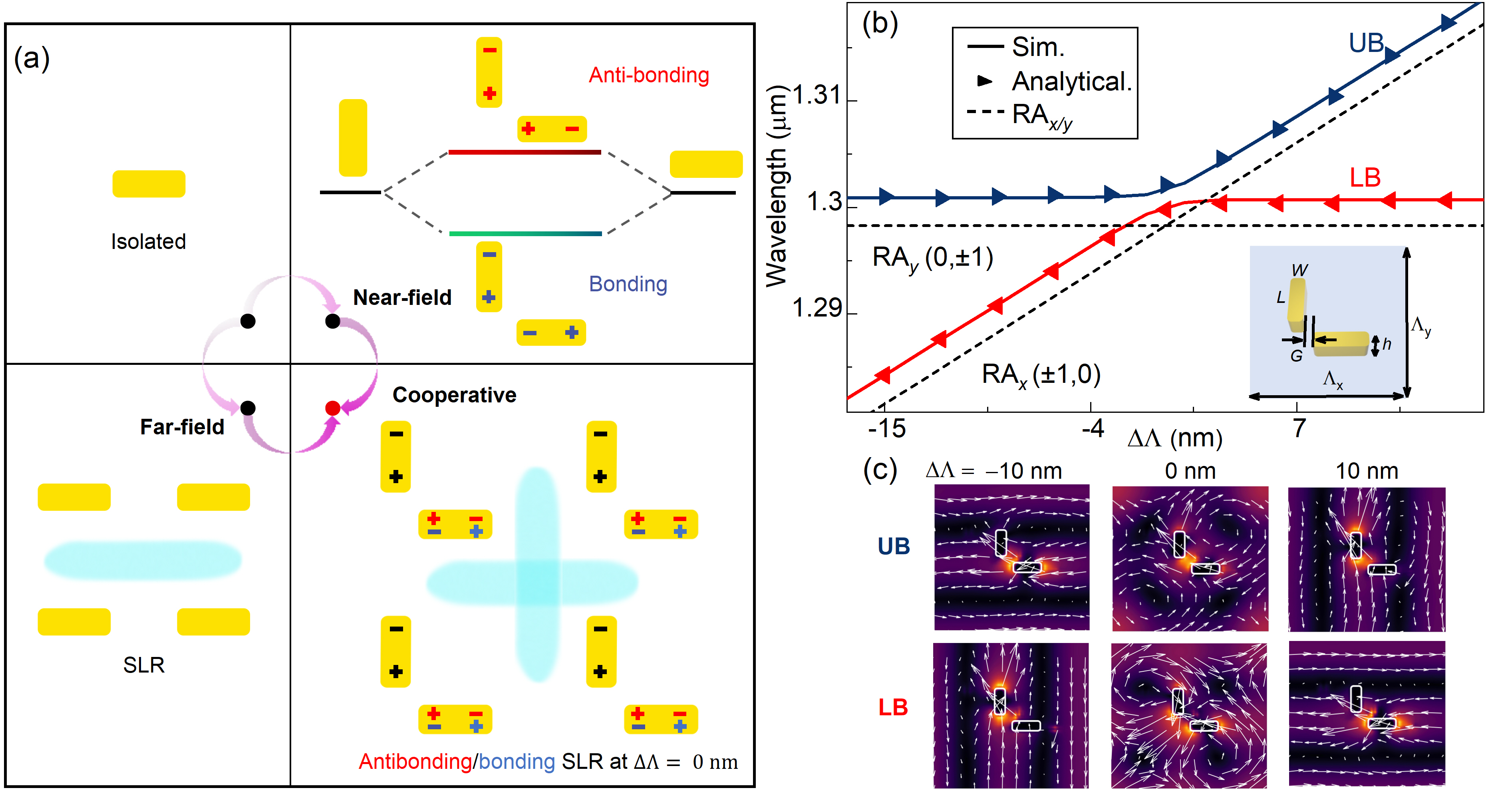}
\caption{Cooperative near- and far-field coupling in periodic L-shaped silver dimers. (a) Schematics of an isolated nanorod, an isolated dimer supporting bonding and anti-bonding LSPRs due to near-field coupling, periodic array of nanorods with SLR due to far-field coupling, and periodic dimers with bonding and anti-bonding SLRs due to cooperative near- and far-field coupling. ``$+$'' and ``$-$'' illustrate near-field coupling induced charge distributions, and cyan cloud indicates far-field coupling via in-plane RA diffraction. (b) Analytical and simulated resonant wavelengths of periodic dimers with different $\Delta \Lambda\equiv \Lambda_x - \Lambda_y$. The dashed lines indicate RA wavelengths of the $(0,\pm 1)$ and $(\pm 1,0)$ orders. (c) Evolution of the near-field electric field distributions (color for intensity and arrows for directions) for the UB and the LB for $\Delta \Lambda = -10$, 0, 10~nm. The calculations were performed with $h = 80$~nm, $L = 150$~nm, $W = 60$~nm, $G = 40$~nm, and $\Delta \Lambda_{y} = 895$~nm.}
\label{fig:schem}
\end{figure*}

Metallic nanoantennas have been shown to significantly influence the behaviors of nearby molecules or emitters thanks to their greatly enhanced near fields, as demonstrated by various surface-enhanced applications including surface-enhanced Raman spectroscopy (SERS), fluorescence, absorption, photovoltaic, nonlinear optics, and sensing, either chiral or not \cite{ChemRev2011Maier_NPReview,JPCM2017Iati_NPReview,AM2020Lei_ChiralNPrev,NanoConv2023Park_NPcReview,CSR2021Wong_ChiralNPrev}. Nanoantennas containing small gaps can amplify the local electromagnetic field intensity by orders of magnitude, resulting in small mode volumes and enhanced light-matter interactions within such ``hotspots'' \cite{ChemRev2011Nordlander_DimerReview,NL2013Halas_DimerRabi,AccChemRes2022Han_HotReview}. Among these configurations, L-shaped metallic dimers composed of two close-packed nanorods, not only have strong field enhancement and light localization, but also can amplify the chiroptical response via the bonding and anti-bonding resonances \cite{NL2013Giessen_LDimerChiral,SciAdv2017LiuN_ChiralPlasReview,AOM2017Fang_ChiralRev,AOM2017Valev_ChiralRev,LSA2020Rho_ChiralNPrev}, which are formed due to the near-field coupling of two localized surface plasmon resonances (LSPRs), as illustrated in Fig.~\ref{fig:schem}(a). However, these resonances suffer from high inherent ohmic and scattering losses, especially in the visible and near-infrared regime.

In order to suppress these losses, far-field coupling is another effective approach that is fundamentally different from the near-field coupling \cite{LPR2013Giessen_PlasStronCoupRev,ChemRev2018Grigorenko_SLRrev}. Surface lattice resonances (SLRs), sometimes referred to as collective resonances, which origin from the coherent far-field coupling between the LSPR and the in-plane Rayleigh anomaly (RA) diffraction wave [Fig.~\ref{fig:schem}(a)], can significantly suppress both the absorption and scattering losses  \cite{ChemRev2018Grigorenko_SLRrev,MatToday2018Odom_SLRrev}. Taking advantages of SLRs, the measured quality factors ($Q$-factors) of plasmonic metasurfaces reach up to 790 in the visible \cite{AOM2023Li_HighQVis} or 2340 in the near-infrared \cite{NC2021Boyd_HighQ}. Recently, bound states in the continuum (BIC) have also been adopted to reduce the radiative loss for plasmonic metasurfaces \cite{LSA2020Zhou_Couple}, and in the mid-infrared the measured $Q$-factor was 180 \cite{SciAdv2022Ren_HighQPlasBIC} and the calculated results can reach 3800 \cite{PRL2024Kivshar_HighQPlasBIC}. Nevertheless, in the near-infrared regime, the maximum calculated $Q$-factor of such plasmonic quasi-BIC reduces to only 627 \cite{PRL2024Kivshar_HighQPlasBIC}. Additionally, higher-$Q$ SLRs \cite{AOM2023Li_HighQVis,NC2021Boyd_HighQ,AOM2019Rivas_HighQ} or quasi-BICs \cite{LSA2020Zhou_Couple,SciAdv2022Ren_HighQPlasBIC,PRL2024Kivshar_HighQPlasBIC} usually accompany with smaller modulation depths, posing challenges in measurements and applications. Therefore, how to significantly suppress the losses of plasmonic resonances while maintaining large modulation depth remains a long-standing open question.

In this Letter, we numerically propose and experimentally demonstrate a strategy to tackle this challenge based on cooperative near- and far-field coupling. This leads to an unreported transition from the weak coupling of LSPRs into strong coupling of nonlocal SLRs even when the gap distance of the dimers varies over a wide range. A generalized lattice sum approximation (LSA) model, showing good agreement with simulation and experimental results, will be developed to reveal the underlying physics. Remarkably, plasmonic resonances with large modulation depths and simulated/measured $Q$-factor of 4050/3120 can be achieved. The critical coupling condition for optimal induced chirality in planar achiral dimers will also be clarified. Extremely high $Q$-factor of 2600/2510 in simulations/measurements, and simulated $g$ factor of 0.47 or measured circular dichroism (CD) of 4\% (equivalent to $g$ factor of 0.37) will be obtained. These measured $Q$-factors exceed the {\sl records} of plasmonic resonances in the visible and near-infrared regimes [see Supplemental Material (SM), S1].

Figure~\ref{fig:schem}(a) illustrates the cooperative near- and far-field coupling in periodic L-shaped silver dimers, of which the unit cell is composed of two orthogonally-packed identical nanorods, as shown in Fig.~\ref{fig:schem}(b) inset. The plasmonic metasurface is under normal incidence of plane wave with electric field $\vec{E}_0 = (E_{0x}, E_{0y})^{T}$. Because of the near-field coupling, the polarizability tensor $\bar{\bar{\alpha}}$ of the L-shaped dimer is not diagonal \cite{ACSN2014Muskens_LDimerChiral}
\begin{equation}
\bar{\bar{\alpha}} = \left( \begin{array}{cc}
    \alpha_{xx} & \alpha_{xy} \\
    \alpha_{yx} & \alpha_{yy} 
\end{array} \right)\,.
    \label{eq:alpha}
\end{equation}
For the L-shaped dimer composing of two identical nanorods, this polarizability tensor can be further simplified with $\alpha_{xx}=\alpha_{yy}=\alpha$ and $\alpha_{yx}=\alpha_{xy}$. On the other hand, chirality can be induced for the achiral dimer due to the anisotropic far-field coupling \cite{AOM2021Govorov_ChiralSLR,LPR2023Li_ChiralSLR}. In order to incorporate both the near- and far-field coupling effects, we extend the standard LSA model, which was developed for periodic nanoparticles of diagonal polarizability tensors \cite{ACR2019Schatz_SLRtheory}, into a generalized model which is also expressed as (see SM S2)
\begin{equation}
    \vec{p} = \bar{\bar{\alpha}} \left( \vec{E}_0  + 
    \bar{\bar{S}} \vec{p}\right) \,,
    \label{eq:LSA}
\end{equation}
where the L-shaped dimer in each unit cell is modeled with a point dipole $\vec{p} = (p_{x}, p_{y})^{T}$, and $\bar{\bar{S}} = {\rm diag} (    S_{xx}, S_{yy} )$ is a diagonal matrix of lattice sums.

By solving Eq.~(\ref{eq:LSA}) we obtain
\begin{equation}
p_{x}  = \frac{\left[ \alpha^{-1} - S_{yy} (1-(\frac{\alpha_{xy}}{\alpha})^2) \right] E_{0x} +  \alpha^{-1} (\frac{\alpha_{xy}}{\alpha}) E_{0y}}{(\alpha^{-1} - S_{+})(\alpha^{-1} - S_{-})} 
  \label{eq:px}
\end{equation}
\begin{equation}
p_{y}  = \frac{\alpha^{-1} (\frac{\alpha_{xy}}{\alpha}) E_{0x} + 
 \left[\alpha^{-1} - S_{xx} (1-(\frac{\alpha_{xy}}{\alpha})^2) \right] E_{0y}}{(\alpha^{-1} - S_{+})(\alpha^{-1} - S_{-})} 
  \label{eq:py}
\end{equation}
where 
\begin{equation}
S_{\pm}  = \frac{S_{xx}+S_{yy}}{2} \pm \sqrt{S_{xx} S_{yy}} \sqrt{\left(\frac{\alpha_{xy}}{\alpha}\right)^2 + \frac{(S_{xx}-S_{yy})^2}{4 S_{xx}S_{yy}}}\,.
  \label{eq:spm}
\end{equation}
Here $(\alpha_{xy}/\alpha)^2$ and $(S_{xx}-S_{yy})^2/(4 S_{xx}S_{yy})$ represent the contributions from the off-diagonal polarizability term and the anisotropic lattice sums due to the near- and far-field coupling, respectively. With Eqs.~(\ref{eq:px}) and (\ref{eq:py}), one can analytically obtain the extinction cross section, the transmittance and the reflectance just like the standard LSA model \cite{ACR2019Schatz_SLRtheory}, and predict resonances with spectral splitting using
\begin{equation}
{\rm Re} \{ \alpha^{-1} \} = {\rm Re} \{ S_{\pm} \} \,.
  \label{eq:ResonancePM}
\end{equation}

Specially, if $\alpha_{xy}=0$, which applies for most nanoparticles such as nanorods, nanodisks, or L-shaped dimers with large enough gaps, one gets $S_{\pm}=S_{xx}$, or $S_{yy}$. In this scenario, the generalized LSA model reduces to the standard model \cite{ACR2019Schatz_SLRtheory}: $p_{x}  = E_{0x}/(\alpha^{-1} - S_{xx}) $, and $p_{y}  = E_{0y}/(\alpha^{-1} - S_{yy})$. In other words, without the off-diagonal term of the polarizability tensor due to the near-field coupling, the two orthogonal SLRs are independent from each other. 

On the other hand, if $S_{xx}=S_{yy}=S$, which can be satisfied when $\Lambda_x=\Lambda_y$, Eq.~(\ref{eq:spm}) reduces to
\begin{equation}
S_{\pm}  = S \left(1 \pm  \frac{\alpha_{xy}}{\alpha}\right)\,.
  \label{eq:spmEqual}
\end{equation}
Compared with the standard LSA model, we find that it is the nontrivial $\alpha_{xy}/\alpha$ that introduces the spectral splitting of the hybridized SLRs. This suggests that the off-diagonal polarizability due to near-field coupling can result in strong coupling of the two orthogonal SLRs at zero detuning. 

In Fig.~\ref{fig:schem}(b) we observe Rabi splitting with an anticrossing in the resonant wavelengths of both the analytical and simulated transmittance spectra (see SM S3 and S4). The upper and the lower branches (UB and LB) locate on the longer wavelength side of the RA lines of the $(\pm 1,0)$ and $(0,\pm 1)$ orders, respectively. This is a typical feature of the SLR. The near-field electric field distributions in Fig.~\ref{fig:schem}(c) show that the UB (or the LB)  evolves from the SLR$_y$ (or the SLR$_x$) for $\Delta \Lambda\lesssim -10$~nm, to the bonding (or the antibonding) SLR at zero detuning ($\Delta \Lambda =0$), as illustrated by Fig.~\ref{fig:schem}(a), and then to the SLR$_x$ (or the SLR$_y$) for $\Delta \Lambda\gtrsim 10$~nm. Here the SLR$_{x/y}$ is associated with the in-plane $(\pm 1,0)$- or $(0,\pm 1)$-order RA diffraction along the $x$ or $y$ direction.

\begin{figure}[!hbt]
\centering
\includegraphics[width=90 mm]{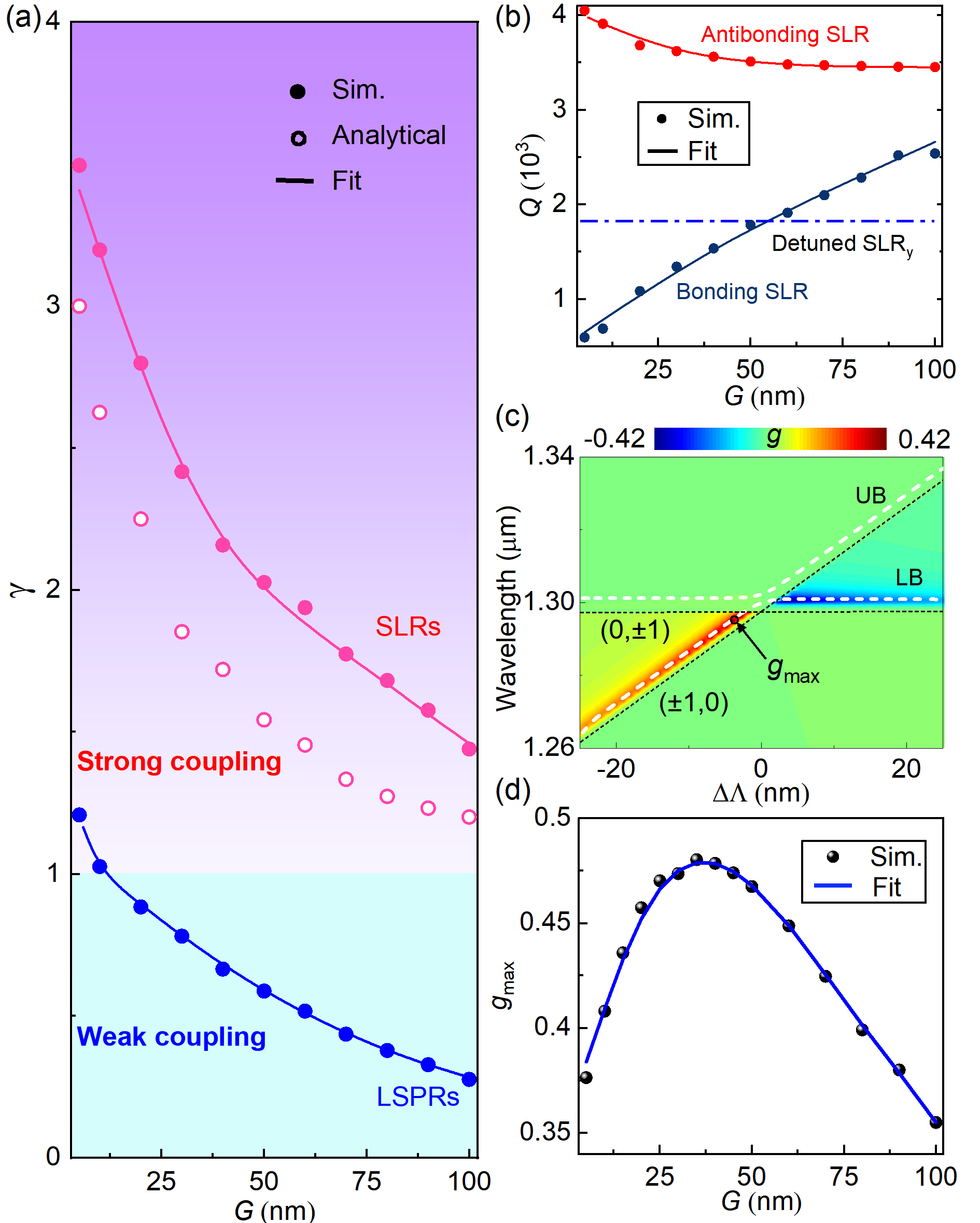}
\caption{Strong coupling, reduced losses, and induced chirility due to cooperative coupling. (a) Coupling strengths for SLRs and LSPRs versus gap distance. The strong and weak coupling regions are bounded by $\gamma=1$. (b) $Q$-factors of antibonding and bonding SLRs versus $G$. The calculations of (a)(b) were performed with $\Delta \Lambda=0$~nm. (c) Simulated $g$ factor spectra as functions of $\Delta \Lambda$ for $G=40$~nm. The UB and LB wavelengths are indicated by gray curves, and the maximum value $g_{\rm max}$ is indicated by a circle. (d) $g_{\rm max}$ as a function of $G$, showing a peak value of 0.48 around $G=35$~nm. }
\label{fig:Gap}
\end{figure}

Under zero detuning, Fig.~\ref{fig:Gap}(a) compares the simulated and analytical coupling strengths of LSPRs and SLRs. Here, the coupling strength is defined as the ratio of the spectral splitting between the two resonances $\Delta \lambda_0$ over the larger linewidth $\delta \lambda_{\rm max} = {\max}\{\delta \lambda_1, \delta \lambda_2 \}$, $\gamma \equiv \Delta \lambda_0 / \delta \lambda_{\rm max}$. For the LSPRs in the isolated L-shaped dimer, the gap distance should be small enough ($G \lesssim 10$~nm) in order to reach the strong coupling region. In contrast, for the SLRs supported by the periodic dimers, the simulated $\gamma$ are larger than 1.5 even when $G$ is as large as 100~nm, meeting the criterion of strong coupling. The analytical $\gamma$, although slightly smaller than the simulated results, also locate in the strong coupling region. In other words, taking advantage of the cooperative near- and far-field coupling, the weak coupling of localized plasmonic resonances is transitioned to the strong coupling of SLRs over a wide range of $G$. 

A consequence of the strong coupling is the great loss redistribution between the antibonding and bonding SLRs with robustness to the gap distance. Corresponding to the stronger coupling strength in Fig.~\ref{fig:Gap}(a) as $G$ decreases, we find in Fig.~\ref{fig:Gap}(b) that the $Q$-factor difference between these two SLRs increases significantly. In other words, stronger coupling exaggerates loss redistribution between the hybridized resonances. Strikingly, for small gap distance of $G=5$~nm, the $Q$-factor of the antibonding SLR reaches up to 4050; even when $G$ is as large as 100~nm, this $Q$-factor remains above 3500, almost twice of that of the detuned SLR$_y$ ($Q=1830$).

\begin{figure*}[!hbt]
\centering
\includegraphics[width=170 mm]{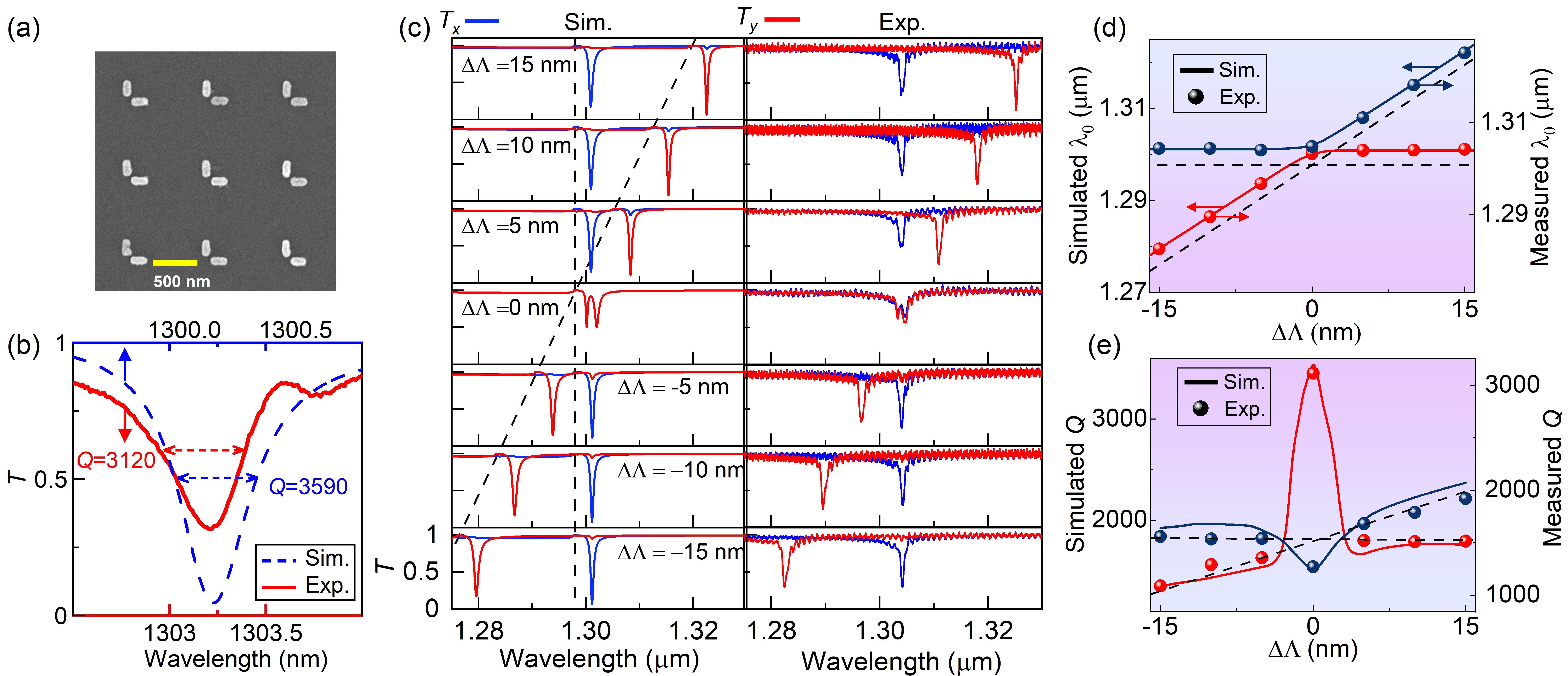}
\caption{Simulated and measured transmittance spectra under linear-polarization. (a) SEM image and (b) transmittance spectra of a typical as-fabricated plasmonic metasurface with $\Delta \Lambda=0$~nm under 45$^{\circ}$ linear-polarization incidence. (c) Transmittance spectra under incidence of $x$-and $y$-polarizations, (d) extracted resonant wavelengths and (e) $Q$-factors for different $\Delta \Lambda$. }
\label{fig:Txy}
\end{figure*}

For the bonding SLR, Fig.~\ref{fig:Gap}(b) shows that the $Q$-factor keeps increasing almost linearly as $G$ increases. Interestingly, when $G\gtrsim 50$~nm, the $Q$-factors of both the antibonding and bonding SLRs are larger than that of the SLR$_y$. Therefore, dual-band hybridized SLRs with $Q$-factors higher than 2000 can be achieved thanks to the cooperative coupling. Such high-$Q$ dual-band resonances can find potential applications such as in enhanced emission \cite{NL2023Wang_dualBandEmisison}, lasing \cite{ACSNano2020Norris_dualBandLasing,AM2021Odom_3BandLasing}, and sensing \cite{LPR2014He_dualBandSens,ACSNano2017Wang_dualBandSens}.


If the lattice sums are anisotropic due to far-field coupling, that is $S_{\rm xx} \neq S_{\rm yy}$, high-$Q$ and strong chiroptical responses can be further induced for the achiral dimer antennas. The chirality can be quantified by Kuhn’s dissymmetry factor $g$, which is defined as preferential absorbance of light of one handedness normalized by total absorbance \cite{Science2011Cohen_SuperChiral,ACR2020Dionne_SuperChiral}, 
\begin{equation}
  g \equiv \frac{2(A_{\rm L}-A_{\rm R})}{A_{\rm L}+A_{\rm R}}\,,
\label{eq:gfactor}
\end{equation}
or alternatively by the CD defined as
\begin{equation}
  {\rm CD} \equiv \frac{T_{\rm L}-T_{\rm R}}{T_{\rm L}+T_{\rm R}}\,,
\label{eq:CD}
\end{equation}
where $A/T_{\rm L/R}$ is the absorbance/transmittance under the normal  incidence of left-/right-hand circular polarization (LCP/RCP). The relationship between the $g$ factor and the CD can be found in SM S4.

Figure~\ref{fig:Gap}(c) shows that as $\Delta \Lambda$ varies, nontrivial $g$ factors with reverse values can be obtained only on the left and right wings of the LB, which are truncated by the RA lines of the $(0,\pm 1)$ and $(\pm 1, 0)$ orders, respectively. Analytical predictions agree well with the simulation results, and the induced chirality should originate from the small absorptance differences between the LCP and RCP incidences (see SM S4). 

The maximum values $g_{\rm max}$ 
can be predicted when
\begin{equation}
  \frac{\partial g}{\partial \Delta \Lambda} =0\,.
\label{eq:gVSdelta}
\end{equation}
By further plotting $g_{\rm max}$ as a function of the gap distance in Fig.~\ref{fig:Gap}(d), we find that $g_{\rm max}$ reaches the peak value of 0.48 when $G\approx 35$~nm, which satisfies
\begin{equation}
\frac{\partial g_{\rm max}}{\partial G} =0\,.
\label{eq:gVSG}
\end{equation}
We refer to Eqs.~(\ref{eq:gVSdelta}) and (\ref{eq:gVSG}) as the critical coupling condition for the optimal induced chirality due to the cooperative coupling effect.

In order to experimentally demonstrate these interesting findings, we fabricated a series of L-shaped silver dimer metasurfaces with $G=40$~nm and varying $\Lambda_x$ (see SM S5). Fig.~\ref{fig:Txy}(a) shows the scanning electron microscope (SEM) image of a typical as-fabricated plasmonic metasurface with $\Delta \Lambda=0$. The simulated and measured transmittance spectra in Fig.~\ref{fig:Txy}(b) show that, the antibonding SLR excited under linear polarization of 45$^\circ$ has large modulation depths of $\sim$100\% and $>$50\%, and meanwhile extremely high $Q$-factors of 3590 and 3120, respectively. This measured $Q$-factor breaks the record for plasmonic resonances in the near infrared regime \cite{NC2021Boyd_HighQ}.

Under different incident polarizations, good agreement between the simulated and measured transmittance spectra can be observed [Fig.~\ref{fig:Txy}(c) and SM S6]. This is better visualized by the resonant wavelengths $\lambda_0$ and the $Q$-factors extracted from these spectra [Fig.~\ref{fig:Txy}(d)(e)]. The deviations between the simulated and measured results should origin from the fabrication imperfections, such as the size/period deviations, and the adoption of a $\sim $0.5~nm chromium wetting layer before depositing the silver film.  

\begin{figure}[!hbt]
\centering
\includegraphics[width=90 mm]{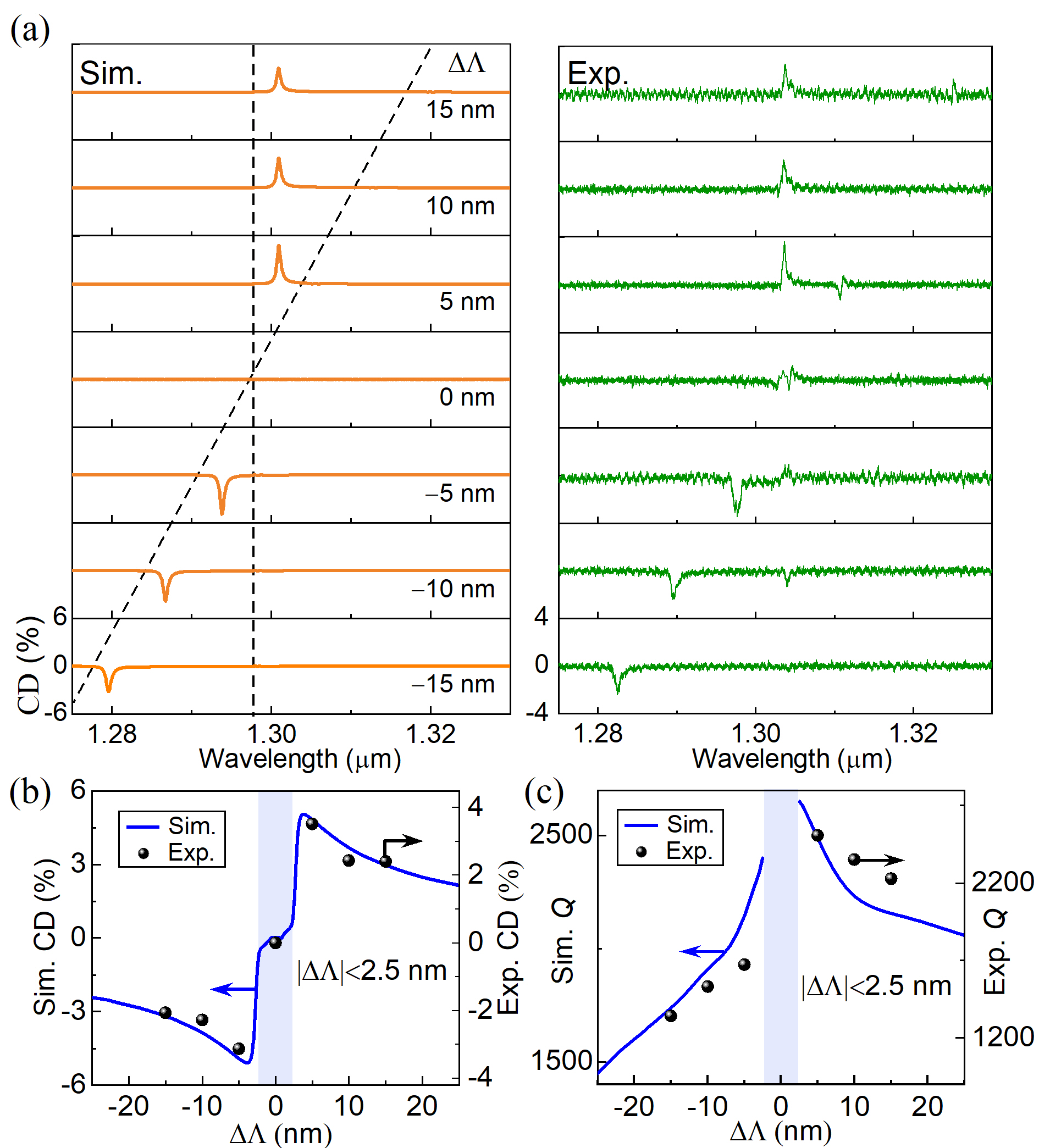}
\caption{Plasmonic chiroptical responses induced by cooperative coupling. (a) Simulated and measured CD spectra, (b) extracted maximum CD values and (c) $Q$-factors for as-fabricated samples with different $\Delta \Lambda$.  Blue regions in (b)(c) indicate rapidly vanishing chirality within $|\Delta \Lambda| < 2.5$~nm.
}
\label{fig:CD}
\end{figure}

With the transmittance spectra for LCP and RCP, we further calculated the CD spectra, and extracted the peak/dip CD values and  $Q$-factors from the CD spectra. Fig.~\ref{fig:CD} shows that the measured data agree with the simulated results. When $\Delta \Lambda=0$~nm, there is no chirality due to mirror symmetry of the periodic array. For $|\Delta \Lambda|<2.5$~nm, the CD values are near zero. As $\Delta \Lambda$ further increases, the simulated CD value first increases dramatically, reaches the maximum around $\Delta \Lambda\sim4$~nm, and then decreases slowly. On the other hand, as $\Delta \Lambda$ increases, the $Q$-factor extracted from the CD spectra also decreases. Strikingly, when $\Delta \Lambda=5$~nm, the simulated/measured maximum CD is 5\%/4\% (the equivalent $g$ factor is 0.47/0.37), and meanwhile the $Q$-factor reaches up to 2600/2510, two orders of magnitude larger than the state of art for planar chiral plasmonic nanoantennas.

In conclusions, we have proposed and demonstrated a new strategy to significantly suppress the loss of plasmonic resonances based on cooperative near- and far-field coupling. A generalized LSA model has been developed to reveal the contributions from the off-diagonal term of the anisotropic polarizability tensor and the anisotropic lattice sums due the near- and far-field coupling effects, respectively. We have observed the transition from the weak near-field coupling between LSPRs in an isolated L-shaped dimer to the strong coupling between SLRs in periodic array with robustness to the gap distance. This results in greatly reduced losses of the antibonding SLR, and high-$Q$ induced chirality for achiral L-shaped dimers under the critical coupling condition. The measured $Q$-factor can reach up to 3120 combined with more than 50\% modulation depth, making the new record for plasmonic resonances in the near-infrared regime. For the chirality induced from the anisotropic periodicity, the measured CD is 4\% (the equivalent $g$ factor is 0.37), and meanwhile the measured $Q$-factor is as high as 2510, two orders of magnitude larger than those of chiral plasmonic nanoantennas in the literature. Our results may solve the long-standing problem of loss in plasmonics and advance the promised applications into practice.

This work was supported by the National Natural Science Foundation of China (62275261, 62405353),
and Natural Science Foundation of Guangdong Province
(2022A1515010086).


\bibliography{myRef.bib}

\providecommand{\noopsort}[1]{}\providecommand{\singleletter}[1]{#1}%
\begin{thebibliography}{33}%
\makeatletter
\providecommand \@ifxundefined [1]{%
 \@ifx{#1\undefined}
}%
\providecommand \@ifnum [1]{%
 \ifnum #1\expandafter \@firstoftwo
 \else \expandafter \@secondoftwo
 \fi
}%
\providecommand \@ifx [1]{%
 \ifx #1\expandafter \@firstoftwo
 \else \expandafter \@secondoftwo
 \fi
}%
\providecommand \natexlab [1]{#1}%
\providecommand \enquote  [1]{``#1''}%
\providecommand \bibnamefont  [1]{#1}%
\providecommand \bibfnamefont [1]{#1}%
\providecommand \citenamefont [1]{#1}%
\providecommand \href@noop [0]{\@secondoftwo}%
\providecommand \href [0]{\begingroup \@sanitize@url \@href}%
\providecommand \@href[1]{\@@startlink{#1}\@@href}%
\providecommand \@@href[1]{\endgroup#1\@@endlink}%
\providecommand \@sanitize@url [0]{\catcode `\\12\catcode `\$12\catcode `\&12\catcode `\#12\catcode `\^12\catcode `\_12\catcode `\%12\relax}%
\providecommand \@@startlink[1]{}%
\providecommand \@@endlink[0]{}%
\providecommand \url  [0]{\begingroup\@sanitize@url \@url }%
\providecommand \@url [1]{\endgroup\@href {#1}{\urlprefix }}%
\providecommand \urlprefix  [0]{URL }%
\providecommand \Eprint [0]{\href }%
\providecommand \doibase [0]{https://doi.org/}%
\providecommand \selectlanguage [0]{\@gobble}%
\providecommand \bibinfo  [0]{\@secondoftwo}%
\providecommand \bibfield  [0]{\@secondoftwo}%
\providecommand \translation [1]{[#1]}%
\providecommand \BibitemOpen [0]{}%
\providecommand \bibitemStop [0]{}%
\providecommand \bibitemNoStop [0]{.\EOS\space}%
\providecommand \EOS [0]{\spacefactor3000\relax}%
\providecommand \BibitemShut  [1]{\csname bibitem#1\endcsname}%
\let\auto@bib@innerbib\@empty
\bibitem [{\citenamefont {Giannini}\ \emph {et~al.}(2011)\citenamefont {Giannini}, \citenamefont {Heck},\ and\ \citenamefont {Maier}}]{ChemRev2011Maier_NPReview}%
  \BibitemOpen
  \bibfield  {author} {\bibinfo {author} {\bibfnamefont {V.}~\bibnamefont {Giannini}}, \bibinfo {author} {\bibfnamefont {A.~I. F.-D. S.~C.}\ \bibnamefont {Heck}},\ and\ \bibinfo {author} {\bibfnamefont {S.~A.}\ \bibnamefont {Maier}},\ }\bibfield  {title} {\bibinfo {title} {Plasmonic nanoantennas: Fundamentals and their use in controlling the radiative properties of nanoemitters},\ }\href@noop {} {\bibfield  {journal} {\bibinfo  {journal} {Chem. Rev.}\ }\textbf {\bibinfo {volume} {111}},\ \bibinfo {pages} {3888} (\bibinfo {year} {2011})}\BibitemShut {NoStop}%
\bibitem [{\citenamefont {Amendola}\ \emph {et~al.}(2017)\citenamefont {Amendola}, \citenamefont {Pilot}, \citenamefont {Frasconi}, \citenamefont {Marag\`o},\ and\ \citenamefont {Iat\`i}}]{JPCM2017Iati_NPReview}%
  \BibitemOpen
  \bibfield  {author} {\bibinfo {author} {\bibfnamefont {V.}~\bibnamefont {Amendola}}, \bibinfo {author} {\bibfnamefont {R.}~\bibnamefont {Pilot}}, \bibinfo {author} {\bibfnamefont {M.}~\bibnamefont {Frasconi}}, \bibinfo {author} {\bibfnamefont {O.~M.}\ \bibnamefont {Marag\`o}},\ and\ \bibinfo {author} {\bibfnamefont {M.~A.}\ \bibnamefont {Iat\`i}},\ }\bibfield  {title} {\bibinfo {title} {Surface plasmon resonance in gold nanoparticles: a review},\ }\href@noop {} {\bibfield  {journal} {\bibinfo  {journal} {J. Phys.: Condens. Matter}\ }\textbf {\bibinfo {volume} {29}},\ \bibinfo {pages} {203002} (\bibinfo {year} {2017})}\BibitemShut {NoStop}%
\bibitem [{\citenamefont {Cao}\ \emph {et~al.}(2020)\citenamefont {Cao}, \citenamefont {Gao}, \citenamefont {Qiu}, \citenamefont {Jin}, \citenamefont {Deng}, \citenamefont {Wong},\ and\ \citenamefont {Lei}}]{AM2020Lei_ChiralNPrev}%
  \BibitemOpen
  \bibfield  {author} {\bibinfo {author} {\bibfnamefont {Z.}~\bibnamefont {Cao}}, \bibinfo {author} {\bibfnamefont {H.}~\bibnamefont {Gao}}, \bibinfo {author} {\bibfnamefont {M.}~\bibnamefont {Qiu}}, \bibinfo {author} {\bibfnamefont {W.}~\bibnamefont {Jin}}, \bibinfo {author} {\bibfnamefont {S.}~\bibnamefont {Deng}}, \bibinfo {author} {\bibfnamefont {K.-Y.}\ \bibnamefont {Wong}},\ and\ \bibinfo {author} {\bibfnamefont {D.}~\bibnamefont {Lei}},\ }\bibfield  {title} {\bibinfo {title} {Chirality transfer from sub-nanometer biochemical molecules to sub-micrometer plasmonic metastructures: Physiochemical mechanisms, biosensing, and bioimaging opportunities},\ }\href@noop {} {\bibfield  {journal} {\bibinfo  {journal} {Adv. Mater.}\ }\textbf {\bibinfo {volume} {32}},\ \bibinfo {pages} {1907151} (\bibinfo {year} {2020})}\BibitemShut {NoStop}%
\bibitem [{\citenamefont {Lee}\ \emph {et~al.}(2023)\citenamefont {Lee}, \citenamefont {Kim},\ and\ \citenamefont {Park}}]{NanoConv2023Park_NPcReview}%
  \BibitemOpen
  \bibfield  {author} {\bibinfo {author} {\bibfnamefont {Y.}~\bibnamefont {Lee}}, \bibinfo {author} {\bibfnamefont {S.}~\bibnamefont {Kim}},\ and\ \bibinfo {author} {\bibfnamefont {J.}~\bibnamefont {Park}},\ }\bibfield  {title} {\bibinfo {title} {Strong coupling in plasmonic metal nanoparticles},\ }\href@noop {} {\bibfield  {journal} {\bibinfo  {journal} {Nano Convergence}\ }\textbf {\bibinfo {volume} {10}},\ \bibinfo {pages} {34} (\bibinfo {year} {2023})}\BibitemShut {NoStop}%
\bibitem [{\citenamefont {Zheng}\ \emph {et~al.}(2021)\citenamefont {Zheng}, \citenamefont {He}, \citenamefont {Kumar}, \citenamefont {Wang}, \citenamefont {Pastoriza-Santos}, \citenamefont {P\'erez-Juste}, \citenamefont {Liz-Marz\'an},\ and\ \citenamefont {Wong}}]{CSR2021Wong_ChiralNPrev}%
  \BibitemOpen
  \bibfield  {author} {\bibinfo {author} {\bibfnamefont {G.}~\bibnamefont {Zheng}}, \bibinfo {author} {\bibfnamefont {J.}~\bibnamefont {He}}, \bibinfo {author} {\bibfnamefont {V.}~\bibnamefont {Kumar}}, \bibinfo {author} {\bibfnamefont {S.}~\bibnamefont {Wang}}, \bibinfo {author} {\bibfnamefont {I.}~\bibnamefont {Pastoriza-Santos}}, \bibinfo {author} {\bibfnamefont {J.}~\bibnamefont {P\'erez-Juste}}, \bibinfo {author} {\bibfnamefont {L.~M.}\ \bibnamefont {Liz-Marz\'an}},\ and\ \bibinfo {author} {\bibfnamefont {K.-Y.}\ \bibnamefont {Wong}},\ }\bibfield  {title} {\bibinfo {title} {Chiral plasmonics},\ }\href@noop {} {\bibfield  {journal} {\bibinfo  {journal} {Chem. Soc. Rev.}\ }\textbf {\bibinfo {volume} {50}},\ \bibinfo {pages} {3738} (\bibinfo {year} {2021})}\BibitemShut {NoStop}%
\bibitem [{\citenamefont {Halas}\ \emph {et~al.}(2011)\citenamefont {Halas}, \citenamefont {Lal}, \citenamefont {Chang}, \citenamefont {Link},\ and\ \citenamefont {Nordlander}}]{ChemRev2011Nordlander_DimerReview}%
  \BibitemOpen
  \bibfield  {author} {\bibinfo {author} {\bibfnamefont {N.~J.}\ \bibnamefont {Halas}}, \bibinfo {author} {\bibfnamefont {S.}~\bibnamefont {Lal}}, \bibinfo {author} {\bibfnamefont {W.-S.}\ \bibnamefont {Chang}}, \bibinfo {author} {\bibfnamefont {S.}~\bibnamefont {Link}},\ and\ \bibinfo {author} {\bibfnamefont {P.}~\bibnamefont {Nordlander}},\ }\bibfield  {title} {\bibinfo {title} {Plasmons in strongly coupled metallic nanostructures},\ }\href@noop {} {\bibfield  {journal} {\bibinfo  {journal} {Chem. Rev.}\ }\textbf {\bibinfo {volume} {111}},\ \bibinfo {pages} {3913} (\bibinfo {year} {2011})}\BibitemShut {NoStop}%
\bibitem [{\citenamefont {Schlather}\ \emph {et~al.}(2013)\citenamefont {Schlather}, \citenamefont {Large}, \citenamefont {Urban}, \citenamefont {Nordlander},\ and\ \citenamefont {Halas}}]{NL2013Halas_DimerRabi}%
  \BibitemOpen
  \bibfield  {author} {\bibinfo {author} {\bibfnamefont {A.~E.}\ \bibnamefont {Schlather}}, \bibinfo {author} {\bibfnamefont {N.}~\bibnamefont {Large}}, \bibinfo {author} {\bibfnamefont {A.~S.}\ \bibnamefont {Urban}}, \bibinfo {author} {\bibfnamefont {P.}~\bibnamefont {Nordlander}},\ and\ \bibinfo {author} {\bibfnamefont {N.~J.}\ \bibnamefont {Halas}},\ }\bibfield  {title} {\bibinfo {title} {Near-field mediated plexcitonic coupling and giant {Rabi} splitting in individual metallic dimers},\ }\href@noop {} {\bibfield  {journal} {\bibinfo  {journal} {Nano Lett.}\ }\textbf {\bibinfo {volume} {13}},\ \bibinfo {pages} {3281} (\bibinfo {year} {2013})}\BibitemShut {NoStop}%
\bibitem [{\citenamefont {Wy}\ \emph {et~al.}(2022)\citenamefont {Wy}, \citenamefont {Jung}, \citenamefont {Hong},\ and\ \citenamefont {Han}}]{AccChemRes2022Han_HotReview}%
  \BibitemOpen
  \bibfield  {author} {\bibinfo {author} {\bibfnamefont {Y.}~\bibnamefont {Wy}}, \bibinfo {author} {\bibfnamefont {H.}~\bibnamefont {Jung}}, \bibinfo {author} {\bibfnamefont {J.~W.}\ \bibnamefont {Hong}},\ and\ \bibinfo {author} {\bibfnamefont {S.~W.}\ \bibnamefont {Han}},\ }\bibfield  {title} {\bibinfo {title} {Exploiting plasmonic hot spots in {Au}-based nanostructures for sensing and photocatalysis},\ }\href@noop {} {\bibfield  {journal} {\bibinfo  {journal} {Acc. Chem. Res.}\ }\textbf {\bibinfo {volume} {55}},\ \bibinfo {pages} {831} (\bibinfo {year} {2022})}\BibitemShut {NoStop}%
\bibitem [{\citenamefont {Yin}\ \emph {et~al.}(2013)\citenamefont {Yin}, \citenamefont {Sch\:aferling}, \citenamefont {Metzger},\ and\ \citenamefont {Giessen}}]{NL2013Giessen_LDimerChiral}%
  \BibitemOpen
  \bibfield  {author} {\bibinfo {author} {\bibfnamefont {X.}~\bibnamefont {Yin}}, \bibinfo {author} {\bibfnamefont {M.}~\bibnamefont {Sch\:aferling}}, \bibinfo {author} {\bibfnamefont {B.}~\bibnamefont {Metzger}},\ and\ \bibinfo {author} {\bibfnamefont {H.}~\bibnamefont {Giessen}},\ }\bibfield  {title} {\bibinfo {title} {Interpreting chiral nanophotonic spectra: The plasmonic {Born-Kuhn} model},\ }\href@noop {} {\bibfield  {journal} {\bibinfo  {journal} {Nano Lett.}\ }\textbf {\bibinfo {volume} {13}},\ \bibinfo {pages} {6238} (\bibinfo {year} {2013})}\BibitemShut {NoStop}%
\bibitem [{\citenamefont {Hentschel}\ \emph {et~al.}(2017)\citenamefont {Hentschel}, \citenamefont {Sch\:aferling}, \citenamefont {Duan}, \citenamefont {Giessen},\ and\ \citenamefont {Liu}}]{SciAdv2017LiuN_ChiralPlasReview}%
  \BibitemOpen
  \bibfield  {author} {\bibinfo {author} {\bibfnamefont {M.}~\bibnamefont {Hentschel}}, \bibinfo {author} {\bibfnamefont {M.}~\bibnamefont {Sch\:aferling}}, \bibinfo {author} {\bibfnamefont {X.}~\bibnamefont {Duan}}, \bibinfo {author} {\bibfnamefont {H.}~\bibnamefont {Giessen}},\ and\ \bibinfo {author} {\bibfnamefont {N.}~\bibnamefont {Liu}},\ }\bibfield  {title} {\bibinfo {title} {Chiral plasmonics},\ }\href@noop {} {\bibfield  {journal} {\bibinfo  {journal} {Sci. Adv.}\ }\textbf {\bibinfo {volume} {3}},\ \bibinfo {pages} {e1602735} (\bibinfo {year} {2017})}\BibitemShut {NoStop}%
\bibitem [{\citenamefont {Luo}\ \emph {et~al.}(2017)\citenamefont {Luo}, \citenamefont {Chi}, \citenamefont {Jiang}, \citenamefont {Li}, \citenamefont {Zu}, \citenamefont {Li},\ and\ \citenamefont {Fang}}]{AOM2017Fang_ChiralRev}%
  \BibitemOpen
  \bibfield  {author} {\bibinfo {author} {\bibfnamefont {Y.}~\bibnamefont {Luo}}, \bibinfo {author} {\bibfnamefont {C.}~\bibnamefont {Chi}}, \bibinfo {author} {\bibfnamefont {M.}~\bibnamefont {Jiang}}, \bibinfo {author} {\bibfnamefont {R.}~\bibnamefont {Li}}, \bibinfo {author} {\bibfnamefont {S.}~\bibnamefont {Zu}}, \bibinfo {author} {\bibfnamefont {Y.}~\bibnamefont {Li}},\ and\ \bibinfo {author} {\bibfnamefont {Z.}~\bibnamefont {Fang}},\ }\bibfield  {title} {\bibinfo {title} {Plasmonic chiral nanostructures: Chiroptical effects and applications},\ }\href@noop {} {\bibfield  {journal} {\bibinfo  {journal} {Adv. Optical Mater.}\ }\textbf {\bibinfo {volume} {5}},\ \bibinfo {pages} {1700040} (\bibinfo {year} {2017})}\BibitemShut {NoStop}%
\bibitem [{\citenamefont {Collins}\ \emph {et~al.}(2017)\citenamefont {Collins}, \citenamefont {Kuppe}, \citenamefont {Hooper}, \citenamefont {Sibilia}, \citenamefont {Centini},\ and\ \citenamefont {Valev}}]{AOM2017Valev_ChiralRev}%
  \BibitemOpen
  \bibfield  {author} {\bibinfo {author} {\bibfnamefont {J.~T.}\ \bibnamefont {Collins}}, \bibinfo {author} {\bibfnamefont {C.}~\bibnamefont {Kuppe}}, \bibinfo {author} {\bibfnamefont {D.~C.}\ \bibnamefont {Hooper}}, \bibinfo {author} {\bibfnamefont {C.}~\bibnamefont {Sibilia}}, \bibinfo {author} {\bibfnamefont {M.}~\bibnamefont {Centini}},\ and\ \bibinfo {author} {\bibfnamefont {V.~K.}\ \bibnamefont {Valev}},\ }\bibfield  {title} {\bibinfo {title} {Chirality and chiroptical effects in metal nanostructures: Fundamentals and current trends},\ }\href@noop {} {\bibfield  {journal} {\bibinfo  {journal} {Adv. Optical Mater.}\ }\textbf {\bibinfo {volume} {5}},\ \bibinfo {pages} {1700182} (\bibinfo {year} {2017})}\BibitemShut {NoStop}%
\bibitem [{\citenamefont {Mun}\ \emph {et~al.}(2020)\citenamefont {Mun}, \citenamefont {Kim}, \citenamefont {Yang}, \citenamefont {Badloe}, \citenamefont {Qiu},\ and\ \citenamefont {Rho}}]{LSA2020Rho_ChiralNPrev}%
  \BibitemOpen
  \bibfield  {author} {\bibinfo {author} {\bibfnamefont {J.}~\bibnamefont {Mun}}, \bibinfo {author} {\bibfnamefont {M.}~\bibnamefont {Kim}}, \bibinfo {author} {\bibfnamefont {Y.}~\bibnamefont {Yang}}, \bibinfo {author} {\bibfnamefont {T.}~\bibnamefont {Badloe}}, \bibinfo {author} {\bibfnamefont {C.-W.}\ \bibnamefont {Qiu}},\ and\ \bibinfo {author} {\bibfnamefont {J.}~\bibnamefont {Rho}},\ }\bibfield  {title} {\bibinfo {title} {Electromagnetic chirality: from fundamentals to nontraditional chiroptical phenomena},\ }\href@noop {} {\bibfield  {journal} {\bibinfo  {journal} {Light: Sci. Appl.}\ }\textbf {\bibinfo {volume} {9}},\ \bibinfo {pages} {139} (\bibinfo {year} {2020})}\BibitemShut {NoStop}%
\bibitem [{\citenamefont {Ameling}\ and\ \citenamefont {Giessen}(2013)}]{LPR2013Giessen_PlasStronCoupRev}%
  \BibitemOpen
  \bibfield  {author} {\bibinfo {author} {\bibfnamefont {R.}~\bibnamefont {Ameling}}\ and\ \bibinfo {author} {\bibfnamefont {H.}~\bibnamefont {Giessen}},\ }\bibfield  {title} {\bibinfo {title} {Microcavity plasmonics: strong coupling of photonic cavities and plasmons},\ }\href@noop {} {\bibfield  {journal} {\bibinfo  {journal} {Laser Photonics Rev.}\ }\textbf {\bibinfo {volume} {7}},\ \bibinfo {pages} {141} (\bibinfo {year} {2013})}\BibitemShut {NoStop}%
\bibitem [{\citenamefont {Kravets}\ \emph {et~al.}(2018)\citenamefont {Kravets}, \citenamefont {Kabashin}, \citenamefont {Barnes},\ and\ \citenamefont {Grigorenko}}]{ChemRev2018Grigorenko_SLRrev}%
  \BibitemOpen
  \bibfield  {author} {\bibinfo {author} {\bibfnamefont {V.~G.}\ \bibnamefont {Kravets}}, \bibinfo {author} {\bibfnamefont {A.~V.}\ \bibnamefont {Kabashin}}, \bibinfo {author} {\bibfnamefont {W.~L.}\ \bibnamefont {Barnes}},\ and\ \bibinfo {author} {\bibfnamefont {A.~N.}\ \bibnamefont {Grigorenko}},\ }\bibfield  {title} {\bibinfo {title} {Plasmonic surface lattice resonances: A review of properties and applications},\ }\href@noop {} {\bibfield  {journal} {\bibinfo  {journal} {Chem. Rev.}\ }\textbf {\bibinfo {volume} {118}},\ \bibinfo {pages} {5912} (\bibinfo {year} {2018})}\BibitemShut {NoStop}%
\bibitem [{\citenamefont {Wang}\ \emph {et~al.}(2018)\citenamefont {Wang}, \citenamefont {Ramezani}, \citenamefont {V{\"a}kev{\"a}inen}, \citenamefont {T{\"o}rm{\"a}}, \citenamefont {Rivas},\ and\ \citenamefont {Odom}}]{MatToday2018Odom_SLRrev}%
  \BibitemOpen
  \bibfield  {author} {\bibinfo {author} {\bibfnamefont {W.}~\bibnamefont {Wang}}, \bibinfo {author} {\bibfnamefont {M.}~\bibnamefont {Ramezani}}, \bibinfo {author} {\bibfnamefont {A.~I.}\ \bibnamefont {V{\"a}kev{\"a}inen}}, \bibinfo {author} {\bibfnamefont {P.}~\bibnamefont {T{\"o}rm{\"a}}}, \bibinfo {author} {\bibfnamefont {J.~G.}\ \bibnamefont {Rivas}},\ and\ \bibinfo {author} {\bibfnamefont {T.~W.}\ \bibnamefont {Odom}},\ }\bibfield  {title} {\bibinfo {title} {The rich photonic world of plasmonic nanoparticle arrays},\ }\href@noop {} {\bibfield  {journal} {\bibinfo  {journal} {Mater. Today}\ }\textbf {\bibinfo {volume} {21}},\ \bibinfo {pages} {303} (\bibinfo {year} {2018})}\BibitemShut {NoStop}%
\bibitem [{\citenamefont {Li}\ \emph {et~al.}(2023)\citenamefont {Li}, \citenamefont {Du}, \citenamefont {Xiong},\ and\ \citenamefont {Yang}}]{AOM2023Li_HighQVis}%
  \BibitemOpen
  \bibfield  {author} {\bibinfo {author} {\bibfnamefont {G.}~\bibnamefont {Li}}, \bibinfo {author} {\bibfnamefont {X.}~\bibnamefont {Du}}, \bibinfo {author} {\bibfnamefont {L.}~\bibnamefont {Xiong}},\ and\ \bibinfo {author} {\bibfnamefont {X.}~\bibnamefont {Yang}},\ }\bibfield  {title} {\bibinfo {title} {Plasmonic metasurfaces with quality factors up to 790 in the visible regime},\ }\href@noop {} {\bibfield  {journal} {\bibinfo  {journal} {Adv. Optical Mater.}\ }\textbf {\bibinfo {volume} {11}},\ \bibinfo {pages} {2301205} (\bibinfo {year} {2023})}\BibitemShut {NoStop}%
\bibitem [{\citenamefont {Bin-Alam}\ \emph {et~al.}(2021)\citenamefont {Bin-Alam}, \citenamefont {Reshef}, \citenamefont {Mamchur}, \citenamefont {Alam}, \citenamefont {Carlow}, \citenamefont {Upham}, \citenamefont {Sullivan}, \citenamefont {M\'enard}, \citenamefont {Huttunen}, \citenamefont {Boyd},\ and\ \citenamefont {Dolgaleva}}]{NC2021Boyd_HighQ}%
  \BibitemOpen
  \bibfield  {author} {\bibinfo {author} {\bibfnamefont {M.~S.}\ \bibnamefont {Bin-Alam}}, \bibinfo {author} {\bibfnamefont {O.}~\bibnamefont {Reshef}}, \bibinfo {author} {\bibfnamefont {Y.}~\bibnamefont {Mamchur}}, \bibinfo {author} {\bibfnamefont {M.~Z.}\ \bibnamefont {Alam}}, \bibinfo {author} {\bibfnamefont {G.}~\bibnamefont {Carlow}}, \bibinfo {author} {\bibfnamefont {J.}~\bibnamefont {Upham}}, \bibinfo {author} {\bibfnamefont {B.~T.}\ \bibnamefont {Sullivan}}, \bibinfo {author} {\bibfnamefont {J.-M.}\ \bibnamefont {M\'enard}}, \bibinfo {author} {\bibfnamefont {M.~J.}\ \bibnamefont {Huttunen}}, \bibinfo {author} {\bibfnamefont {R.~W.}\ \bibnamefont {Boyd}},\ and\ \bibinfo {author} {\bibfnamefont {K.}~\bibnamefont {Dolgaleva}},\ }\bibfield  {title} {\bibinfo {title} {Ultra-high-{$Q$} resonances in plasmonic metasurfaces},\ }\href@noop {} {\bibfield  {journal} {\bibinfo  {journal} {Nat. Commun.}\ }\textbf {\bibinfo {volume} {12}},\ \bibinfo {pages} {974} (\bibinfo {year} {2021})}\BibitemShut {NoStop}%
\bibitem [{\citenamefont {Lin}\ \emph {et~al.}(2020)\citenamefont {Lin}, \citenamefont {Qiu}, \citenamefont {Zhang}, \citenamefont {Guo}, \citenamefont {Cai}, \citenamefont {Xiao}, \citenamefont {He},\ and\ \citenamefont {Zhou}}]{LSA2020Zhou_Couple}%
  \BibitemOpen
  \bibfield  {author} {\bibinfo {author} {\bibfnamefont {J.}~\bibnamefont {Lin}}, \bibinfo {author} {\bibfnamefont {M.}~\bibnamefont {Qiu}}, \bibinfo {author} {\bibfnamefont {X.}~\bibnamefont {Zhang}}, \bibinfo {author} {\bibfnamefont {H.}~\bibnamefont {Guo}}, \bibinfo {author} {\bibfnamefont {Q.}~\bibnamefont {Cai}}, \bibinfo {author} {\bibfnamefont {S.}~\bibnamefont {Xiao}}, \bibinfo {author} {\bibfnamefont {Q.}~\bibnamefont {He}},\ and\ \bibinfo {author} {\bibfnamefont {L.}~\bibnamefont {Zhou}},\ }\bibfield  {title} {\bibinfo {title} {Tailoring the lineshapes of coupled plasmonic systems based on a theory derived from first principles},\ }\href@noop {} {\bibfield  {journal} {\bibinfo  {journal} {Light: Sci. Appl.}\ }\textbf {\bibinfo {volume} {9}},\ \bibinfo {pages} {158} (\bibinfo {year} {2020})}\BibitemShut {NoStop}%
\bibitem [{\citenamefont {Aigner}\ \emph {et~al.}(2022)\citenamefont {Aigner}, \citenamefont {Tittl}, \citenamefont {Wang}, \citenamefont {Weber}, \citenamefont {Kivshar}, \citenamefont {Maier},\ and\ \citenamefont {Ren}}]{SciAdv2022Ren_HighQPlasBIC}%
  \BibitemOpen
  \bibfield  {author} {\bibinfo {author} {\bibfnamefont {A.}~\bibnamefont {Aigner}}, \bibinfo {author} {\bibfnamefont {A.}~\bibnamefont {Tittl}}, \bibinfo {author} {\bibfnamefont {J.}~\bibnamefont {Wang}}, \bibinfo {author} {\bibfnamefont {T.}~\bibnamefont {Weber}}, \bibinfo {author} {\bibfnamefont {Y.}~\bibnamefont {Kivshar}}, \bibinfo {author} {\bibfnamefont {S.~A.}\ \bibnamefont {Maier}},\ and\ \bibinfo {author} {\bibfnamefont {H.}~\bibnamefont {Ren}},\ }\bibfield  {title} {\bibinfo {title} {Plasmonic bound states in the continuum to tailor light-matter coupling},\ }\href@noop {} {\bibfield  {journal} {\bibinfo  {journal} {Phys. Rev. Lett.}\ }\textbf {\bibinfo {volume} {8}},\ \bibinfo {pages} {eadd4816} (\bibinfo {year} {2022})}\BibitemShut {NoStop}%
\bibitem [{\citenamefont {Liang}\ \emph {et~al.}(2024)\citenamefont {Liang}, \citenamefont {Tsai},\ and\ \citenamefont {Kivshar}}]{PRL2024Kivshar_HighQPlasBIC}%
  \BibitemOpen
  \bibfield  {author} {\bibinfo {author} {\bibfnamefont {Y.}~\bibnamefont {Liang}}, \bibinfo {author} {\bibfnamefont {D.~P.}\ \bibnamefont {Tsai}},\ and\ \bibinfo {author} {\bibfnamefont {Y.}~\bibnamefont {Kivshar}},\ }\bibfield  {title} {\bibinfo {title} {From local to nonlocal high-{$Q$} plasmonic metasurfaces},\ }\href@noop {} {\bibfield  {journal} {\bibinfo  {journal} {Phys. Rev. Lett.}\ }\textbf {\bibinfo {volume} {133}},\ \bibinfo {pages} {053801} (\bibinfo {year} {2024})}\BibitemShut {NoStop}%
\bibitem [{\citenamefont {Le-Van}\ \emph {et~al.}(2019)\citenamefont {Le-Van}, \citenamefont {Zoethout}, \citenamefont {Geluk}, \citenamefont {Ramezani}, \citenamefont {Berghuis},\ and\ \citenamefont {Rivas}}]{AOM2019Rivas_HighQ}%
  \BibitemOpen
  \bibfield  {author} {\bibinfo {author} {\bibfnamefont {Q.}~\bibnamefont {Le-Van}}, \bibinfo {author} {\bibfnamefont {E.}~\bibnamefont {Zoethout}}, \bibinfo {author} {\bibfnamefont {E.-J.}\ \bibnamefont {Geluk}}, \bibinfo {author} {\bibfnamefont {M.}~\bibnamefont {Ramezani}}, \bibinfo {author} {\bibfnamefont {M.}~\bibnamefont {Berghuis}},\ and\ \bibinfo {author} {\bibfnamefont {J.~G.}\ \bibnamefont {Rivas}},\ }\bibfield  {title} {\bibinfo {title} {Enhanced quality factors of surface lattice resonances in plasmonic arrays of nanoparticles},\ }\href@noop {} {\bibfield  {journal} {\bibinfo  {journal} {Adv. Optical Mater.}\ }\textbf {\bibinfo {volume} {7}},\ \bibinfo {pages} {1801451} (\bibinfo {year} {2019})}\BibitemShut {NoStop}%
\bibitem [{\citenamefont {Black}\ \emph {et~al.}(2014)\citenamefont {Black}, \citenamefont {Wang}, \citenamefont {de~Groot}, \citenamefont {Arbouet},\ and\ \citenamefont {Muskens}}]{ACSN2014Muskens_LDimerChiral}%
  \BibitemOpen
  \bibfield  {author} {\bibinfo {author} {\bibfnamefont {L.-J.}\ \bibnamefont {Black}}, \bibinfo {author} {\bibfnamefont {Y.}~\bibnamefont {Wang}}, \bibinfo {author} {\bibfnamefont {C.~H.}\ \bibnamefont {de~Groot}}, \bibinfo {author} {\bibfnamefont {A.}~\bibnamefont {Arbouet}},\ and\ \bibinfo {author} {\bibfnamefont {O.~L.}\ \bibnamefont {Muskens}},\ }\bibfield  {title} {\bibinfo {title} {Optimal polarization conversion in coupled dimer plasmonic nanoantennas for metasurfaces},\ }\href@noop {} {\bibfield  {journal} {\bibinfo  {journal} {ACS Nano}\ }\textbf {\bibinfo {volume} {13}},\ \bibinfo {pages} {6390} (\bibinfo {year} {2014})}\BibitemShut {NoStop}%
\bibitem [{\citenamefont {Movsesyan}\ \emph {et~al.}(2021)\citenamefont {Movsesyan}, \citenamefont {Besteiro}, \citenamefont {Kong}, \citenamefont {Wang},\ and\ \citenamefont {Govorov}}]{AOM2021Govorov_ChiralSLR}%
  \BibitemOpen
  \bibfield  {author} {\bibinfo {author} {\bibfnamefont {A.}~\bibnamefont {Movsesyan}}, \bibinfo {author} {\bibfnamefont {L.~V.}\ \bibnamefont {Besteiro}}, \bibinfo {author} {\bibfnamefont {X.-T.}\ \bibnamefont {Kong}}, \bibinfo {author} {\bibfnamefont {Z.}~\bibnamefont {Wang}},\ and\ \bibinfo {author} {\bibfnamefont {A.~O.}\ \bibnamefont {Govorov}},\ }\bibfield  {title} {\bibinfo {title} {Engineering strongly chiral plasmonic lattices with achiral unit cells for sensing and photodetection},\ }\href@noop {} {\bibfield  {journal} {\bibinfo  {journal} {Adv. Optical Mater.}\ }\textbf {\bibinfo {volume} {9}},\ \bibinfo {pages} {2101943} (\bibinfo {year} {2021})}\BibitemShut {NoStop}%
\bibitem [{\citenamefont {Luo}\ \emph {et~al.}(2023)\citenamefont {Luo}, \citenamefont {Du}, \citenamefont {Huang},\ and\ \citenamefont {Li}}]{LPR2023Li_ChiralSLR}%
  \BibitemOpen
  \bibfield  {author} {\bibinfo {author} {\bibfnamefont {X.}~\bibnamefont {Luo}}, \bibinfo {author} {\bibfnamefont {X.}~\bibnamefont {Du}}, \bibinfo {author} {\bibfnamefont {R.}~\bibnamefont {Huang}},\ and\ \bibinfo {author} {\bibfnamefont {G.}~\bibnamefont {Li}},\ }\bibfield  {title} {\bibinfo {title} {High-{$Q$} and strong chiroptical responses in planar metasurfaces empowered by mie surface lattice resonances},\ }\href@noop {} {\bibfield  {journal} {\bibinfo  {journal} {Laser Photonics Rev.}\ }\textbf {\bibinfo {volume} {17}},\ \bibinfo {pages} {2300186} (\bibinfo {year} {2023})}\BibitemShut {NoStop}%
\bibitem [{\citenamefont {Cherqui}\ \emph {et~al.}(2019)\citenamefont {Cherqui}, \citenamefont {Bourgeois}, \citenamefont {Wang},\ and\ \citenamefont {Schatz}}]{ACR2019Schatz_SLRtheory}%
  \BibitemOpen
  \bibfield  {author} {\bibinfo {author} {\bibfnamefont {C.}~\bibnamefont {Cherqui}}, \bibinfo {author} {\bibfnamefont {M.~R.}\ \bibnamefont {Bourgeois}}, \bibinfo {author} {\bibfnamefont {D.}~\bibnamefont {Wang}},\ and\ \bibinfo {author} {\bibfnamefont {G.~C.}\ \bibnamefont {Schatz}},\ }\bibfield  {title} {\bibinfo {title} {Plasmonic surface lattice resonances: Theory and computation},\ }\href@noop {} {\bibfield  {journal} {\bibinfo  {journal} {Acc. Chem. Res.}\ }\textbf {\bibinfo {volume} {52}},\ \bibinfo {pages} {2548} (\bibinfo {year} {2019})}\BibitemShut {NoStop}%
\bibitem [{\citenamefont {Zhang}\ \emph {et~al.}(2023)\citenamefont {Zhang}, \citenamefont {Xu}, \citenamefont {Liu}, \citenamefont {Lang}, \citenamefont {Zhang}, \citenamefont {Li}, \citenamefont {Lu}, \citenamefont {Chen}, \citenamefont {Wang}, \citenamefont {Wang},\ and\ \citenamefont {Li}}]{NL2023Wang_dualBandEmisison}%
  \BibitemOpen
  \bibfield  {author} {\bibinfo {author} {\bibfnamefont {Z.}~\bibnamefont {Zhang}}, \bibinfo {author} {\bibfnamefont {C.}~\bibnamefont {Xu}}, \bibinfo {author} {\bibfnamefont {C.}~\bibnamefont {Liu}}, \bibinfo {author} {\bibfnamefont {M.}~\bibnamefont {Lang}}, \bibinfo {author} {\bibfnamefont {Y.}~\bibnamefont {Zhang}}, \bibinfo {author} {\bibfnamefont {M.}~\bibnamefont {Li}}, \bibinfo {author} {\bibfnamefont {W.}~\bibnamefont {Lu}}, \bibinfo {author} {\bibfnamefont {Z.}~\bibnamefont {Chen}}, \bibinfo {author} {\bibfnamefont {C.}~\bibnamefont {Wang}}, \bibinfo {author} {\bibfnamefont {S.}~\bibnamefont {Wang}},\ and\ \bibinfo {author} {\bibfnamefont {X.}~\bibnamefont {Li}},\ }\bibfield  {title} {\bibinfo {title} {Dual control of enhanced quasi-bound states in the continuum emission from resonant c‑{Si} metasurfaces},\ }\href@noop {} {\bibfield  {journal} {\bibinfo  {journal} {Nano Lett.}\ }\textbf {\bibinfo {volume} {23}},\ \bibinfo {pages} {7584} (\bibinfo {year} {2023})}\BibitemShut {NoStop}%
\bibitem [{\citenamefont {Winkler}\ \emph {et~al.}(2020)\citenamefont {Winkler}, \citenamefont {Ruckriegel}, \citenamefont {Henar~Rojo}, \citenamefont {Leo}, \citenamefont {Rabouw},\ and\ \citenamefont {Norris}}]{ACSNano2020Norris_dualBandLasing}%
  \BibitemOpen
  \bibfield  {author} {\bibinfo {author} {\bibfnamefont {J.~M.}\ \bibnamefont {Winkler}}, \bibinfo {author} {\bibfnamefont {M.~J.}\ \bibnamefont {Ruckriegel}}, \bibinfo {author} {\bibfnamefont {R.~C.~K.}\ \bibnamefont {Henar~Rojo}}, \bibinfo {author} {\bibfnamefont {E.~D.}\ \bibnamefont {Leo}}, \bibinfo {author} {\bibfnamefont {F.~T.}\ \bibnamefont {Rabouw}},\ and\ \bibinfo {author} {\bibfnamefont {D.~J.}\ \bibnamefont {Norris}},\ }\bibfield  {title} {\bibinfo {title} {Dual-wavelength lasing in quantum-dot plasmonic lattice lasers},\ }\href@noop {} {\bibfield  {journal} {\bibinfo  {journal} {ACS Nano}\ }\textbf {\bibinfo {volume} {14}},\ \bibinfo {pages} {5223} (\bibinfo {year} {2020})}\BibitemShut {NoStop}%
\bibitem [{\citenamefont {Guan}\ \emph {et~al.}(2021)\citenamefont {Guan}, \citenamefont {Li}, \citenamefont {Juarez}, \citenamefont {Sample}, \citenamefont {Wang}, \citenamefont {Schatz},\ and\ \citenamefont {Odom}}]{AM2021Odom_3BandLasing}%
  \BibitemOpen
  \bibfield  {author} {\bibinfo {author} {\bibfnamefont {J.}~\bibnamefont {Guan}}, \bibinfo {author} {\bibfnamefont {R.}~\bibnamefont {Li}}, \bibinfo {author} {\bibfnamefont {X.~G.}\ \bibnamefont {Juarez}}, \bibinfo {author} {\bibfnamefont {A.~D.}\ \bibnamefont {Sample}}, \bibinfo {author} {\bibfnamefont {Y.}~\bibnamefont {Wang}}, \bibinfo {author} {\bibfnamefont {G.~C.}\ \bibnamefont {Schatz}},\ and\ \bibinfo {author} {\bibfnamefont {T.~W.}\ \bibnamefont {Odom}},\ }\bibfield  {title} {\bibinfo {title} {Plasmonic nanoparticle lattice devices for white-light lasing},\ }\href@noop {} {\bibfield  {journal} {\bibinfo  {journal} {Adv. Mater.}\ }\textbf {\bibinfo {volume} {35}},\ \bibinfo {pages} {2103262} (\bibinfo {year} {2021})}\BibitemShut {NoStop}%
\bibitem [{\citenamefont {Lin}\ \emph {et~al.}(2014)\citenamefont {Lin}, \citenamefont {Zhang}, \citenamefont {Qian},\ and\ \citenamefont {He}}]{LPR2014He_dualBandSens}%
  \BibitemOpen
  \bibfield  {author} {\bibinfo {author} {\bibfnamefont {J.}~\bibnamefont {Lin}}, \bibinfo {author} {\bibfnamefont {Y.}~\bibnamefont {Zhang}}, \bibinfo {author} {\bibfnamefont {J.}~\bibnamefont {Qian}},\ and\ \bibinfo {author} {\bibfnamefont {S.}~\bibnamefont {He}},\ }\bibfield  {title} {\bibinfo {title} {A nano-plasmonic chip for simultaneous sensing with dual-resonance surface-enhanced raman scattering and localized surface plasmon resonance},\ }\href@noop {} {\bibfield  {journal} {\bibinfo  {journal} {Laser Photonics Rev.}\ }\textbf {\bibinfo {volume} {8}},\ \bibinfo {pages} {610} (\bibinfo {year} {2014})}\BibitemShut {NoStop}%
\bibitem [{\citenamefont {Chen}\ \emph {et~al.}(2017)\citenamefont {Chen}, \citenamefont {Wang}, \citenamefont {Yao},\ and\ \citenamefont {Wang}}]{ACSNano2017Wang_dualBandSens}%
  \BibitemOpen
  \bibfield  {author} {\bibinfo {author} {\bibfnamefont {X.}~\bibnamefont {Chen}}, \bibinfo {author} {\bibfnamefont {C.}~\bibnamefont {Wang}}, \bibinfo {author} {\bibfnamefont {Y.}~\bibnamefont {Yao}},\ and\ \bibinfo {author} {\bibfnamefont {C.}~\bibnamefont {Wang}},\ }\bibfield  {title} {\bibinfo {title} {Plasmonic vertically coupled complementary antennas for dual-mode infrared molecule sensing},\ }\href@noop {} {\bibfield  {journal} {\bibinfo  {journal} {ACS Nano}\ }\textbf {\bibinfo {volume} {11}},\ \bibinfo {pages} {8034} (\bibinfo {year} {2017})}\BibitemShut {NoStop}%
\bibitem [{\citenamefont {Tang}\ and\ \citenamefont {Cohen}(2011)}]{Science2011Cohen_SuperChiral}%
  \BibitemOpen
  \bibfield  {author} {\bibinfo {author} {\bibfnamefont {Y.}~\bibnamefont {Tang}}\ and\ \bibinfo {author} {\bibfnamefont {A.~E.}\ \bibnamefont {Cohen}},\ }\bibfield  {title} {\bibinfo {title} {Enhanced enantioselectivity in excitation of chiral molecules by superchiral light},\ }\href@noop {} {\bibfield  {journal} {\bibinfo  {journal} {Science}\ }\textbf {\bibinfo {volume} {332}},\ \bibinfo {pages} {333} (\bibinfo {year} {2011})}\BibitemShut {NoStop}%
\bibitem [{\citenamefont {Solomon}\ \emph {et~al.}(2020)\citenamefont {Solomon}, \citenamefont {Saleh}, \citenamefont {Tadesse},\ and\ \citenamefont {Dionne}}]{ACR2020Dionne_SuperChiral}%
  \BibitemOpen
  \bibfield  {author} {\bibinfo {author} {\bibfnamefont {M.~L.}\ \bibnamefont {Solomon}}, \bibinfo {author} {\bibfnamefont {A.~A.~E.}\ \bibnamefont {Saleh}}, \bibinfo {author} {\bibfnamefont {L.~F.}\ \bibnamefont {Tadesse}},\ and\ \bibinfo {author} {\bibfnamefont {J.~A.}\ \bibnamefont {Dionne}},\ }\bibfield  {title} {\bibinfo {title} {Nanophotonic platforms for chiral sensing and separation},\ }\href@noop {} {\bibfield  {journal} {\bibinfo  {journal} {Acc. Chem. Res.}\ }\textbf {\bibinfo {volume} {53}},\ \bibinfo {pages} {588} (\bibinfo {year} {2020})}\BibitemShut {NoStop}%
\end{thebibliography}%

\end{document}